\documentclass[12pt]{iopart}
\usepackage{graphicx}
\usepackage{cite}

\begin{document}
 
\title[Quantum dynamics of an optomechanical system in the presence]{Quantum dynamics of an optomechanical system in the presence of photonic Bose-Einstein condensate}

\author{M Fani$^1$ and M H Naderi$^2$}

\address{$^1$Department of Physics, Faculty of Science, University of Isfahan, Hezar Jerib, 81746-73441, Isfahan, Iran}
\address{$^2$Quantum Optics Group, Department of Physics, Faculty of Science, University of Isfahan, Hezar Jerib, 81746-73441, Isfahan, Iran}
\ead{mhnaderi@sci.ui.ac.ir}
\vspace{10pt}
\begin{indented}
\item[]October 2016
\end{indented}

\begin{abstract}
In this paper, we study theoretically the optomechanical interaction of an almost pure condensate of photons with an oscillating mechanical membrane in a micro-cavity. We show that in the Bogoliubov approximation, due to the large number of photons in the condensate phase, there is a linear strong effective coupling between the Bogoliubov mode of the photonic Bose-Einstein condensate (BEC) and the mechanical motion of the membrane which depends on the nonlinear photon-photon scattering potential. This coupling leads to the cooling of the mechanical motion, the normal mode splitting (NMS), the squeezing of the output field and the entanglement between the excited mode of the cavity and the mechanical mode. We show that, in one hand, the nonlinearity of the photon gas increases the degree of the squeezing of the output field of the micro-cavity and the efficiency of the cooling process at high temperatures. In the other hand, it reduces NMS in the displacement spectrum of the oscillating membrane and the degree of the optomechanical entanglement. In addition, the temperature of the photonic BEC can be used to control the above-mentioned phenomena.
\end{abstract}

%
\vspace{2pc}
%
%
\maketitle
%
%

\section{Introduction}

The interactions between light and matter have been studied and implemented in a wide variety of systems from cavity QED to solid-state systems for many years \cite{Auffeves}. One type of light-matter interactions which has attracted much attention over the past decade is the optomechanical(OM) coupling between radiation pressure and a mechanical oscillator (for a recent review, see, e.g., \cite{Aspelmeyer}). In an OM cavity, the electromagnetic field affects the mechanical motion of a movable mirror via radiation pressure resulting the OM coupling between the cavity field and the mechanical element. Because of the dependence of the cavity length on the intensity of the field, the OM interaction is intrinsically nonlinear \cite{Gong}. The OM coupling was first considered to trap and control the dielectric particles \cite{Ashkin} and then to detect the gravitational waves \cite{Braginsky}. However, the field of optomechanics has undergone a rapid progress during the past years and is currently subject to intensive research investigations. Nowadays, the OM coupling can be realized in many different configurations with a wide range of mechanical frequencies (from $kHz$ to a few $GHz$) and of effective masses (from $pg$ to $kg$) \cite{Aspelmeyer}. 

As a few examples of OM setups, we can mention to the Fabry-perot cavities with a moving end mirror \cite{Metzger}, suspended dielectric membranes \cite{Thompson,Sankey}, photonic crystal cavities \cite{Eichenfield,Gavartin} and cold atoms trapped inside optical cavities \cite{Brennecke,Murch}. Furthermore, different applications have been considered for the OM systems such as high precision detection and measurement of small forces, displacements and masses \cite{Aspelmeyer}. Nevertheless, the feature that makes the OM system more interesting is the fact that it is one of the most promising candidates for exploring quantum effects in the mesoscopic and macroscopic scales \cite{Schliesser,Brooks}. 
The main obstacle to observe quantum behaviour in the macroscopic scales comes from thermal noise. As a matter of fact, in order to prepare a macroscopic mechanical object in a quantum state it is necessary to cool it down to its motional ground state. In the OM setups, due to the finite lifetime of the photons, the radiation pressure force is non-conservative and it can provide an extra mechanical damping under certain circumstances which leads to the cooling of the mechanical element \cite{Aspelmeyer}. This cooling mechanism, which is referred to as back-action (or self) cooling in the literature \cite{Genes2}, has been studied theoretically \cite{Dantan,Genes3,Dobrindt} and it was experimentally realized in the regime of a few phonons \cite{Schliesser,Groblacher} and even close to the ground state \cite{Chan}. The possibility of the ground state cooling has been also predicted theoretically in the resolved side band regime \cite{Genes3,Marquardt} but it has not yet been achieved experimentally.  

Optomechanical systems have attracted considerable attention in connection with their ability to generate entangled states of macroscopic objects. The radiation pressure-induced entanglement between two mirrors of a ring cavity was first proposed in \cite{Mancini}. After that, many other schemes have been proposed to generate entanglement between different subsytems in the standard OM as well as hybrid OM systems \cite{Vitali,Tian,Vitali2,Akram}.  Aside from the generation of entanglement, the possibility of producing nonclassical sates of both the mechanical motion and cavity field have been investigated in various OM configurations \cite{Clerk,Purdy,Safavi-Naeini}. One other noticeable feature of the OM systems is relevant to the phenomenon of normal mode splitting (NMS) which stems from the strong coupling of two degenerate modes with energy exchange taking place on a time scale faster than the decoherence of each mode \cite{Dobrindt}. The optomechanical NMS, which has been experimentally observed \cite{Teufel}, may be taken into account in those experiments that seek to demonstrate ground-state cooling of the mechanical oscillator \cite{Marquardt,Wilson-Rae}. In recent years, there has been an increasing interest in nonlinear hybrid OM cavities, where the nonlinearity is mainly contributed by the nonlinear media, such as optical Kerr medium \cite{Kumar}, optical parametric amplifier(OPA) \cite{Huang2}, or combination of both (Kerr-down conversion nonlinearity) \cite{Shahidani}. It has been shown that the Kerr nonlinearity shifts the cavity frequency and weakens the OM coupling \cite{Kumar}, the OPA leads to strong OM coupling via increasing the intensity of the cavity field \cite{Huang2}, and the Kerr-down conversion can lead to significant photon-phonon entanglement simultaneously with ground-state cooling of the oscillating mirror \cite{Shahidani}.

On the other hand, hybrid optical cavities containing an atomic Bose-Einstein condensate(BEC) have been identified as suitable candidates for realization of the OM coupling arising from the dispersive interaction of the BEC with the cavity field \cite{Brennecke2}. In such systems, the fluctuations of the atomic field, i.e., the Bogoliubov mode, plays the role of the vibrational mode of the mechanical oscillator in an OM cavity \cite{Stamper-Kurn}. In comparison to the standard OM setups, the OM cavities assisted by BEC operate in a different regime and provides a OM strong coupling as well as the Kerr nonlinearity in the low-photon number regime \cite{Gupta}. Furthermore, it has been shown \cite{Dalafi} that in the Bogoliubov approximation the atomic collision affects the dynamics of the system by shifting the energy of the excited mode and provides an atomic parametric amplifier interaction.

Despite the bosonic nature of photons, it was believed for a long time that the realization of \textit{photonic} BEC faces a fundamental obstacle. The problem lies in vanishing mass and  chemical potential of photons which make it very difficult to cool a fixed number of photons such that they form a condensate. However, the BEC phase transition was observed experimentally for the first time at room temperature for photons trapped in a dye-filled optical microcavity with two curved mirrors at a small distance in comparison to their dimensions \cite{Klaers2,Klaers1}. 
The cavity mirrors provide a non-vanishing mass in the paraxial approximation \cite{Chiao} as well as a harmonic trapping potential for photons \cite{Klaers2,Klaers1}. In addition, photons can be thermalized via multiple absorption and re-emission by the dye molecules \cite{Klaers1}. In this way, the photon gas is formally equivalent to a two-dimensional trapped massive boson gas and so the photonic BEC phase transition is possible. It should be noted that in contrast to the laser, the condensation of photons is a thermal equilibrium phase transition.

Following the first successful experimental realization, the BEC of photons, as a new state of light, has attracted much attention in recent years. Besides further relevant experimental investigations \cite{Schmitt1,Schmitt2,Marelic}, various theoretical studies have been carried out to explain the equilibration and properties of the system in the framework of statistical mechanics \cite{Sobyanin,Klaers3,Weiss}, non equilibrium Green's function \cite{deLeeuw,deLeeuw2,Vanderwurff} and nonlinear Schr\"odinger equation \cite{Nyman,Strinati}. In addition, some other theoretical schemes have been proposed to achieve thermalization and BEC phase transition of a photon gas in a dilute non-degenerated atomic gas \cite{Kruchkov}, in a one-dimensional barrel optical microresonator filled with a dye solution \cite{Cheng}, in an optomechanical cavity with a segmented moving mirror \cite{Weitz}, and in a multimode hybrid atom-membrane optomechanical microcavity \cite{Fani}. Furthermore, the temperature-dependent decay rate \cite{Zhang} and enhanced dynamic stark shift \cite{Fan} of an atom interacting with a BEC of photons have been theoretically studied. 

In spite of various works which have been done on the photonic BEC, the issue of its interaction with matter systems have received less attention. Inspired by the existing studies on the interaction between atomic BEC and OM systems and also motivated by the similarities between photonic and atomic BECs, in this paper we consider the linear OM coupling of a BEC of photons with the mechanical oscillator in an OM cavity. We study the dynamical effects of the radiation pressure force induced by the photonic BEC. The cooling of the mechanical mode and the steady-state entanglement between the mechanical mode and the Bogoliubov mode of the BEC are investigated. The coherence of the photonic condensate causes an effective strong coupling between the collective excitations of the condensate and the mechanical modes. In addition, the evidences of a weak photon-photon interaction in the photonic BEC has been observed in the experiment \cite{Klaers2}. Actually, this interaction has significant influence on the behaviour of the OM system, so it can provide a tool to extract information about the photonic BEC via measuring the properties of the OM system. We show that the photon-photon interaction and BEC temperature can also be considered as the control parameters to achieve quantum effects.

The remainder of this paper is organized as follows. In the section \ref{secmodel}, we introduce the physical model of the system under consideration. In section \ref{secbogo}, by considering an almost pure condensate with weak two-body interaction, we apply the Bogoliubov approximation to the Hamiltonian of the system. We devote section \ref{secmotion} to the derivation the equations of motion describing the system dynamics within the input-output formalism. In section \ref{secdis}, we study the cooling and the displacement spectrum of the mechanical mode. Section \ref{secinten} is allocated to investigate the output intensity and quadrature squeezing spectra and in section \ref{secentan}, the steady-state entanglement between the collective excitation mode of the photonic BEC and the mechanical mode is studied. Finally, we summarize our conclusions in section \ref{seccon}.

\section{Physical Model}\label{secmodel}

\begin{figure}[htbp]
\centering
\includegraphics[width=3in]{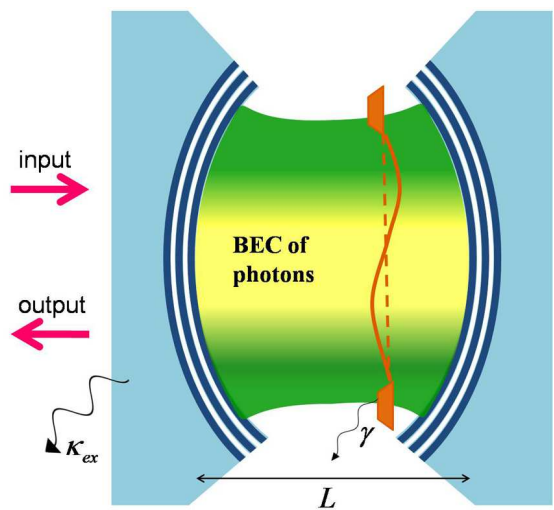}
\caption{(Color online) Schematic diagram of an optomechanical microcavity of length $L$ composed of a vibrating membrane with damping rate $\gamma$ in the presence of a BEC of photons. The cavity is coupled to the external environment via one of its mirrors with decay rate $\kappa _{ex} $. Furthermore, the cavity field is driven with an input coherent field.}
\label{fig1}
\end{figure}

In this section we model the optomechanical interaction of an oscillating micromechanical element with a BEC of photons in a microcavity which is pumped by a coherent field (figure \ref{fig1}). The microcavity consists of two curved mirrors with curvature $R$ which are fixed at distance $L$ from each other. When $L \ll R$ the longitudinal mode number can be fixed and the cavity field is equivalent to an effectively massive two-dimensional photon gas which, due to the mirrors curvature, is confined in a harmonic trap with frequency $\omega_{t}=\frac{c}{\sqrt{R L/2}} $ where $c$ is the speed of light \cite{Klaers2,Klaers1}. Therefore, the frequency of each cavity mode in the paraxial approximation (i. e., the longitudinal wave number $k_{\parallel}$ to be much larger than the transverse wave number $k_{\perp}$) is given by \cite{Rexin} $\omega_{k} = c | \textbf{k} |$ where 
\begin{equation}\label{k}
\omega_{k}  \simeq c({k_\parallel } + \frac{1}{2}\frac{{{k_ \bot }^2}}{{{k_\parallel }}} )= \frac{{2\pi n c}}{L} + ({2l + \left| m \right| + 1}) \omega _t,
\end{equation}
with $l$ and $m$ being, respectively, the radial and the azimuthal quantum numbers. Thus, the transverse wave number is given by ${k_ \bot } = \frac{{2\pi n}}{L}\frac{\textit{s}}{{\sqrt {RL/2} }}$ where for the sake of brevity, we have defined $\textit{s}=2l + \left| m \right| +1$ which is also the degeneracy of the cavity modes. In addition, the fixed longitudinal mode number $n$ determines the cut-off frequency of the cavity, $\omega_{cut} = 2n\pi c/ L$, as well as the effective mass of photons, $m_{photon}=\hbar \omega_{cut} / c^2 $. The photon gas can be in thermal equilibrium with a reservoir at temperature $T$ with non-zero chemical potential and so it can undergo a BEC phase transition when the total number of photons in the cavity $N_t \simeq \frac{2 L P}{\hbar \omega_{cut} c}  $ (with $P$ the pump power) is larger than the critical photon number $N_c \simeq \frac{\pi^2}{3} (\frac{k_B T}{\hbar \omega_{t}})^2$\cite{Klaers2}.
In the experiment of photon BEC generation \cite{Klaers1} the phase transition occurred at the critical power of $(1.55 \pm 0.6)W$ corresponding to the critical photon number $N_c = (6.3 \pm 2.4)\times 10^4$ at room temperature ($k_B T/\hbar \omega_t \simeq 150$). In that experiment, the effective mass of photons and the trap frequency are $m_{photon}\simeq 6.7 \times 10^{-36} kg$ and $\omega _t \simeq 2\pi \times 4.1 \times 10^{10} Hz$, respectively.

As shown schematically in figure \ref{fig1}, the mechanical element is considered to be a semi-transparent, thin dielectric membrane inside the cavity which can vibrate at eigenfrequencies $\Omega_{q} = v_s \left| \textbf{q} \right| $, where $v_s$ is the sound velocity and $\textbf{q}$ is the associated two-dimensional wave vector of phonons. 
The Hamiltonian of the system in the grand canonical ensemble can be written as 
 \numparts
\begin{eqnarray}
 H = {H_f} + {H_m} + {H_I};\label{H1a}\\\nonumber
{H_f} = \sum\limits_k {(\hbar {\delta _k} - \mu )a_k^\dagger {a_k}} \\
\,\,\,\,\,\,+ \frac{1}{2}\sum\limits_q {V(q)a_{k + q}^\dagger a_{k' - q}^\dagger {a_{k'}}{a_k}}
 + i\sum\limits_k {{\eta _k}(a_k^\dagger  - {a_k})} ,\label{H1b}\\
{H_m} = \sum\limits_q {\hbar {\Omega _q}c_q^\dagger {c_q}} ,\label{H1c}\\
{H_I} =  - \sum\limits_k {\hbar {g_k}(q)a_{k + q}^\dagger {a_k}({c_q} + c_{-q}^\dagger )} \label{H1d} ,
\end{eqnarray}
\endnumparts
where all the summations are taken over two-dimensional wave vectors. The term $H_f$ is the Hamiltonian of the photon gas with chemical potential $\mu$ and $a_k$($a_k^{\dagger}$) is the annihilation(creation) operator of a photon in the mode $\textbf{k}$. The second term in the Hamiltonian $H_f$ denotes the photon-photon scattering with the interaction potential $V(q)$, which can arise from the Kerr nonlinearity or thermal lensing in the cavity medium \cite{Klaers2}. The third term in the Hamiltonian of (\ref{H1b}) describes the cavity pumping by a coherent field with real amplitude $\eta_k$ and frequency $\omega_p$. The Hamiltonian $H_f$  has been written in the frame rotating with pump frequency, so $\delta_k = \omega_k - \omega_p$. Equation (\ref{H1c}) is the free Hamiltonian of the mechanical oscillator where $c_k$($c_k^{\dagger}$) is the annihilation(creation) operator of a phonon in the mode $\textbf{q}$. The Hamiltonian of (\ref{H1d}) represents the interaction of the membrane motion and the cavity field, where $g_k(q)$ is the linear OM coupling constant. This part of the total Hamiltonian describes the scattering of photons between the different cavity modes by the mechanical motion of the membrane. In the interaction Hamiltonian, due to the large free spectral range of the cavity field, we have assumed that the membrane oscillations can not induce transitions between the modes with different longitudinal modes. 

In order to have a non-vanishing chemical potential and so a BEC of photons, it is necessary to have a thermalization mechanism in the micro-cavity, such as multiple absorption and reemission of photons by a dye solution \cite{Klaers2,Klaers1} or other proposed thermalization schemes \cite{Weitz,Fani,Hafezi}. Here, we have assumed that the thermalization process takes place much faster than the optomechanical coupling, which is relevant for the typical experiment data \cite{Klaers1,Aspelmeyer}. Thus, during the optomechanical interaction the photon gas remains in the thermal equilibrium, so the thermalization mechanism has not been included in the Hamiltonian of the system and it is not also shown in the figure \ref{fig1}. However, we will take into account the effects of the thermalization process as a thermal reservoir for the cavity field (see section \ref{secmotion}). In addition, the cavity field is assumed to be coupled to an external reservoir via one of the end mirrors and we consider a vacuum input noise due to this coupling. It is also worth pointing out that, depending on the type of the microcavity, other kinds of mechanical elements (instead of the micromechanical membrane) might be considered such as photonic crystal membranes \cite {Gavartin}, surface acoustic waves on a mirror \cite{Shi}, segmented \cite{Weitz}, or deformable \cite{Antoni} end mirror . In this paper, we analyse our results based on the experimentally feasible parameters for the case of micromechanical membrane setup, where the mechanical frequency is of the order of $MHz$, while the single-photon OM coupling is of the order of $kHz$ \cite{Sankey,Eichenfield}. Also the cavity decay rate is set to be a few $MHz$.

\section{Bogoliubov Approximation}\label{secbogo}

To proceed further, we now assume that the total number of photons is far above the BEC threshold, $N_t \gg N_c$, so there is a macroscopic number of photons ($N_0 \gg 1 $) in the ground state of the cavity field and the chemical potential is equal to the lowest energy of the cavity field. On the other hand, since the two-body interaction is weak in this system \cite{Klaers2} we can apply the Bogoliubov theory to the photon gas \cite{Chiao,Zhang}. Replacing $a_0$,  $a^{\dagger}_0$ ($l,m=0$) by $\sqrt{N_0}$ and keeping only leading terms in $\sqrt{N_0}$ in each term of the Hamiltonians of equations (\ref{H1b}) and (\ref{H1d}), we obtain  
 \numparts
\begin{eqnarray}
{H_f} \simeq {E_0} + \sum\limits_{k \ne 0} {(\hbar {\delta _k} - \mu )a_k^\dagger {a_k}} \nonumber \\
\,\, + \frac{1}{2}{V_0} N_0 \sum\limits_{k \ne 0} {(a_{- k}^\dagger a_k^\dagger  + {a_{-k}}{a_k})}  + i\sum\limits_{k \ne 0} {{\eta _k}(a_k^\dagger  - {a_k})} ,\label{H2a}\\
{H_I} \simeq  - \sum\limits_{q \ne 0} {2 \hbar g(q) \sqrt{N_0}(a_q^\dagger  + {a_{ - q}})({c_q} + c_{ - q}^\dagger )} ,\label{H2b}
\end{eqnarray}
\endnumparts
 where $E_0$ is a constant and does not affect the dynamics of the system. Here, the indices $k$ and $q$ denote only the transverse wave numbers ($k_{\perp}$), so the summations are taken over $ \{ lm \} \neq \{ 00 \} $. In equations (\ref{H2a},\ref{H2b}) we have neglected the dependence of OM coupling on $k$, which is justified in the paraxial region.
 Equation (\ref{H2a}) shows that in the Bogoliubov approximation the photon-photon interaction takes the form of a parametric coupling between the modes $\textbf{k} $ and $-\textbf{k}$ which can lead to photon squeezing. We have also assumed that the two-body interaction potential is constant \cite{Zhang} and it is given by $V_0=\zeta (\hbar \omega_{trap})/ 2\pi$ where $\zeta$ is the dimensionless interaction parameter whose experimental value is $\zeta \approx (7\pm 3)\times 10^{-4}$  \cite {Klaers2}. On the other hand, the Bogoliubov approximation leads to the linearization of the OM interaction Hamiltonian of equation (\ref{H2b}). It means that we have considered only the scattering of photons from/into the condensate phase by the OM coupling and the scattering among excited modes has been neglected due to the small population of these modes.

Here, the important feature is that the coherence of the photonic BEC, i. e., the macroscopic ground state population of the photon gas not only provides a strong effective linear coupling among the excited modes of the cavity and the mechanical modes, but also it causes a strong parametric interaction. The condensate photon number is given by $N_0 = N_t - \sum\limits_{k \ne 0} { \langle a_k^\dagger {a_k} \rangle } $ but as we will see in section \ref{secdis}, the changes in the value of $N_0$ induced by the OM coupling and the photon-photon interaction is negligible. Therefore, in the numerical calculation we use the value of $N_0$ for noninteracting photon gas, $N_0 = N_t - \sum\limits_{k \ne 0} { (e^{\hbar (\omega_k - \mu)/k_B T}-1)^{-1} } $.

It is well-known \cite{pitaevskii} that the Bogoliubov transformation,
\begin{eqnarray}\label{bogo}
{a_k} = {u_k}{b_k} + {v_k}b_{ - k}^\dagger , \nonumber \\ 
a_{k}^\dagger  = {u_k}b_k^\dagger  + {v_k}{b_{ - k}},
\end{eqnarray}
with
\begin{eqnarray}\label{uv}
{u_k} = {[\frac{1}{2}(1 + \frac{{{N_0}{V_0}}}{{\hbar {\omega _k}}}){(1 + 2\frac{{{N_0}{V_0}}}{{\hbar {\omega _k}}})^{ - 1/2}} + \frac{1}{2}]^{1/2}},\nonumber \\
{v_k} =  - {[\frac{1}{2}(1 + \frac{{{N_0}{V_0}}}{{\hbar {\omega _k}}}){(1 + 2\frac{{{N_0}{V_0}}}{{\hbar {\omega _k}}})^{ - 1/2}} - \frac{1}{2}]^{1/2}},
\end{eqnarray}
diagonalizes the interaction term in the Hamiltonian $H_f$. Therefore, the total Hamiltonian of the system in terms of the operators $b_k, b^{\dagger}_k$ is given by 
\begin{eqnarray}
H \simeq \sum\limits_{k \ne 0} {\hbar {\tilde{\delta} _k}b_k^\dagger {b_k}}  +\sum\limits_{k\neq 0} {{\Omega _k}c_k^\dagger {c_k}} \nonumber \\
\,\,\, - \sum\limits_{k \ne 0} {\hbar {{\bar g}_k}(b_k^\dagger  + {b_{ - k}})({c_k} + c_{ - k}^\dagger )}  + i\sum\limits_{k \ne 0} {{{\bar \eta }_k}(b_k^\dagger  - {b_k})} ,
\end{eqnarray}
where the detuning is $\tilde{\delta}_k = \tilde{\omega}_k - \omega _p$, with the Bogoliubov dispersion relation ${{\tilde \omega }_k} = {\omega _k}{[1 + 2{N_0}{V_0}/\hbar {\omega _k}]}^{1/2}$. In addition, we have introduced the effective coupling strength $\bar g_k = 2 \sqrt{N_0} g(k)(u_k + v_k)$ and the real effective pump amplitude $\bar \eta _k = \eta _k (v_k - u_k)$. Increasing the interaction parameter $V_0$ causes $\bar{g_k}$ to decrease. Besides, for fixed value of $N_t$, increasing the temperature $T$ reduces $N_0$ slightly, but since $N_0\gg 1$, the effective coupling does not change by $T$, considerably.

In order to more simplify the Hamiltonian, we first apply the unitary transformation $U = \exp \{ i\sum\limits_k {{{\bar \eta }_k}(b_k^\dagger  + {b_k})} /{{\tilde \delta }_k}\} $, which yields ${b_k} \to {b_k} + i{{\bar \eta }_k}/{{\tilde \delta }_k}$ and so the elimination of the coherent pump term. Then by introducing the symmetric and antisymmetric modes,
\numparts
\begin{eqnarray}\label{sym}
B_k^{s/a} = ({b_k} \pm {b_{ - k}})/\sqrt 2 ,\\
C_k^{s/a} = ({c_k} \pm {c_{ - k}})/\sqrt 2 ,
\end{eqnarray}
\endnumparts
the Hamiltonian can be written as
\numparts
\begin{eqnarray}\label{Hsaa}
H &= {H_s} + {H_a};\\
{H_s} &= \sum\limits_{k > 0} {\hbar {\delta _k}B{{_k^s}^\dagger }B_k^s}  + \sum\limits_{k > 0} {\hbar {\Omega _k}C{{_k^s}^\dagger }C_k^s} \nonumber \\
 &- \sum\limits_{k > 0} {\hbar {{\bar g}_k}(B_k^s + B{{_k^s}^\dagger })(C_k^s + C{{_k^s}^\dagger })} ,\label{Hsab}\\
{H_a} &= \sum\limits_{k > 0} {\hbar {\delta _k}B{{_k^a}^\dagger }B_k^a}  + \sum\limits_{k > 0} {\hbar {\Omega _k}C{{_k^a}^\dagger }C_k^a} \nonumber \\
& - \sum\limits_{k > 0} {\hbar {{\bar g}_k}(B_k^a - B{{_k^a}^\dagger })(C_k^a - C{{_k^a}^\dagger })} ,\label{Hsac}
\end{eqnarray}
\endnumparts
where the constant terms have been omitted. In the Hamiltonian of the equation (\ref{Hsaa}) the symmetric and antisymmetric modes are completely decoupled. On the other hand, because of the similarities of the Hamiltonians $H_s$ and $H_a$ most of the physical results are similar to each other for these modes. Thus, in the following sections we consider only the symmetric modes and, for the sake of simplicity, we omit the index $s$ in the corresponding relations. In addition, It is obvious from the Hamiltonian of the equations (\ref{Hsab},\ref{Hsac}) that all the modes $k$ are decoupled and their evolutions are independent, so we also omit the index $k$. Therefore, in the following the symbol $Y$ is used instead of $Y^s_k$  where $Y$ is any of the operators or parameters of the system. 
On the other hand, by adjusting the position and the geometry of the membrane it is possible to select one of the normal modes of the membrane to interact effectively with the cavity field and the coupling of other modes can be neglected \cite{Biancofiore}. In the following, the numerical results are only presented for the mode given by $\textit{s}=2$, which is corresponding to the first excited mode of the cavity.

\section{Dynamics of The System}\label{secmotion}

To describe the dynamics of the system, we apply the input-output formalism \cite{Clerk2}. Using the Hamiltonian of equation (\ref{Hsab}), we can write the equations of motion as follows
\numparts
\begin{eqnarray}\label{motiona}
\dot B &=  - (i\tilde \delta  + \kappa )B + i\bar g(C + {C^\dagger }) \nonumber \\
 &+ \sqrt {2\kappa_{ex} } {B_{in}}+ \sqrt {2\kappa_{0} } F,\\
\dot C &=  - (i\Omega  + \gamma )C + i\bar g(B + {B^\dagger }) + {\xi _c},\label{motionb}
\end{eqnarray}
\endnumparts
with $\kappa = \kappa_{ex} + \kappa_0$ being the total cavity decay rate, where $\kappa_{ex}$ is the cavity decay rate at the input mirror and $\kappa_0$ is the remaining loss rate. The damping rate of the membrane is denoted by $\gamma$ and the operator $B_{in}$ is the input noise of the symmetric Bogoliubov mode. Considering a vacuum input noise for the cavity field, $\langle a_{in} (t) a^{\dagger}_{in} (t') \rangle = \delta (t-t')$ and using equation (\ref{bogo}) and (\ref{sym}) we have 
\begin{equation}\label{bnoiset}
\begin{array}{l}
\langle {B_{in}^\dagger (t){B_{in}}(t')} \rangle  = {v^2}\delta (t - t'),{\mkern 1mu} {\kern 1pt} {\mkern 1mu} \\ \langle {{B_{in}}(t)B_{in}^\dagger (t')} \rangle  = {u^2}\delta (t - t'),\\
\langle {{B_{in}}(t){B_{in}}(t')} \rangle  =  - uv\delta (t - t').
\end{array}
\end{equation}
In addition, the thermalization process is considered as coupling to a thermal reservoir with associated noise operator $F$ whose nonzero correlation functions are given by, $\langle {F^\dagger (t){F}(t')} \rangle  = {{\bar n}_{th}}\delta (t - t'), \langle {{F}(t)F^\dagger (t')} \rangle  = ({{\bar n}_{th}} + 1)\delta (t - t')$  where ${\bar n}_{th} = (e^{\hbar \tilde{\omega}/k_B T}-1)^{-1}$. Therefore, the total noise operator for the Bogoliubov mode can be defined as $\xi_B =\sqrt {2\kappa_{ex} } {B_{in}}+ \sqrt {2\kappa_{0} } F $. Similarly, the mechanical element is in equilibrium with a reservoir at temperature $T_m$, so the noise operator $\xi _c$ satisfies $\langle {\xi _c^\dagger  (t){\xi _c}(t')} \rangle  = 2 \gamma {{\bar n}_c}\delta (t - t'), \langle {{\xi _c}(t)\xi _c^\dagger (t')} \rangle  =2 \gamma ({{\bar n}_c} + 1)\delta (t - t')$ in which $\bar n_c=(e^{\hbar \Omega / k_B T_m}-1)^{-1}$. It should be noticed that equation (\ref{motionb}) is valid when $\Omega \gg \gamma$ which is justified for most OM systems \cite{Aspelmeyer}. 

Defining the dimensionless quadratures
\numparts
\begin{eqnarray}\label{quadraturs}
X = \frac{1}{{\sqrt 2 }}(B + {B^\dagger }),P = \frac{1}{{\sqrt 2 i}}(B - {B^\dagger }),\\
x = \frac{1}{{\sqrt 2 }}(C + {C^\dagger }),p = \frac{1}{{\sqrt 2 i}}(C - {C^\dagger }),
\end{eqnarray}
\endnumparts
the equations of motion (\ref{motiona}) and (\ref{motionb}) can be written in the compact matrix form
\begin{equation}\label{matrixeq}
{\bf{\dot u}}(t) = {\bf{Au}}(t) + {\bf{n}}(t),
\end{equation}
with ${\bf{u}}(t) = {(X(t),P(t),x(t),p(t))^T}$ and the corresponding noise operator vector is ${\bf{n}}(t) = ({\xi _X, \xi _P, \xi _x, \xi _p)^T}$ where we have defined $\xi _X = (\xi_B+\xi_B^{\dagger})/\sqrt{2}$ , $\xi _P =-i (\xi_B-\xi_B^{\dagger})/\sqrt{2}$, $\xi_x =(\xi_c+\xi_c^{\dagger})/\sqrt{2}$, $\xi_p =-i (\xi_c-\xi_c^{\dagger})/\sqrt{2} $. The drift matrix ${\bf{A}}$ is given by
\begin{equation}\label{drift}
{\bf{A}} = \left( {\begin{array}{*{20}{c}}
{ - \kappa }&{\tilde \delta }&0&0\\
{ - \tilde \delta }&{ - \kappa }&{2\bar g}&0\\
0&0&{ - \gamma }&\Omega \\
{2\bar g}&0&{ - \Omega }&{ - \gamma }
\end{array}} \right).
\end{equation}

The system is stable when the real part of all the eigenvalues of the drift matrix $\bf{A}$ are negative. The stability condition can be checked by the Routh-Hurwitz criterion \cite{Gradshteyn}. For the stable system with Gaussian noises all the stationary properties of the system can be extracted from the Lyapunov equation \cite{Genes}
\begin{equation}\label{lyap}
{\bf{AV}} + {\bf{V}}{{\bf{A}}^{\bf{T}}} =  - {\bf{D}},
\end{equation}
where $\textbf{V}$ is the $4 \times 4$ stationary covariance matrix with components ${V_{ij}} = \frac{1}{2} \langle {{u_i}(\infty ){u_j}(\infty ) + {u_j}(\infty ){u_i}(\infty )} \rangle $ and the corresponding diffusion matrix $\bf D$ is given by ${\bf{D}} = Diag[\{ \kappa_{ex} (u-v)^2 +\kappa_0 (2{{\bar n}_{th}} + 1),\kappa_{ex} (u+v)^2 +\kappa_0 (2{{\bar n}_{th}} + 1),\gamma (2{{\bar n}_c} + 1),\gamma (2{{\bar n}_c} + 1)\} ]$. It should be noted that the covariance matrix $\bf{V}$ which is obtained from equation (\ref{lyap}) is associated to the symmetric Bogoliubov mode $B$. Using the Bogoliubov transformation of equation (\ref{bogo}), we obtain the following covariance matrix for the symmetric cavity mode, $A_k = (a_k + a_{-k})/\sqrt{2}$  in terms of the elements of $\bf{V}$
\begin{equation}\label{Vp}
\resizebox{0.65\textwidth}{!}{${{\bf{V'}} = \left( {\begin{array}{*{20}{c}}
{{{(u + v)}^2}{V_{11}}}&{({u^2} - {v^2}){V_{12}}}&{(u + v){V_{13}}}&{(u + v){V_{14}}}\\
{({u^2} - {v^2}){V_{21}}}&{{{(u - v)}^2}{V_{22}}}&{(u - v){V_{23}}}&{(u - v){V_{24}}}\\
{(u + v){V_{31}}}&{(u - v){V_{32}}}&{{V_{33}}}&{{V_{34}}}\\
{(u + v){V_{41}}}&{(u - v){V_{42}}}&{{V_{43}}}&{{V_{44}}}
\end{array}} \right).}$}
\end{equation}

\section{Displacement spectrum and cooling of the membrane}\label{secdis} 

To study the dynamical effects of the optomechanical coupling on the mechanical oscillator in the presence of the Bogoliubov modes of a photonic BEC, we drive the mechanical susceptibility and the displacement spectrum of the oscillator. The symmetrized displacement spectrum of the mechanical oscillator is defined as \cite{Genes3} 
\begin{equation}\label{dispspedef}
\resizebox{0.65\textwidth}{!}{${S_x}(\omega ) = \frac{1}{{4\pi }}\int {d\omega '{e^{ - i(\omega  + \omega ')t}} \langle {x(\omega )x(\omega ') + x(\omega ')x(\omega )} \rangle } ,$}
\end{equation}
where $x(\omega)$, the Fourier transformation of $x(t)$, can be obtained by solving the time-domain equation of motion (\ref{matrixeq}) in the frequency domain. The Fourier transform of the time-domain operator $f(\tau)$ is defined by$f(\omega)=\frac{1}{\sqrt{2\pi}} \int_{ - \infty }^\infty  {d\tau f(\tau ){e^{ - i\omega \tau }}} $. In this manner, one obtains $x(\omega)=\chi_m(\omega) F_m(\omega)$ where the mechanical susceptibility $\chi _m (\omega)$ is given by
\begin{equation}\label{ki}
\chi _m (\omega ) = \frac{\Omega }{{{\Omega _{eff}}^2 - {\omega ^2} - i\omega {\gamma _{eff}}}},
\end{equation}
with the effective mechanical frequency,
\begin{equation}\label{omegaeff}
{\Omega _{eff}}^2 = {\gamma ^2} + {\Omega ^2} - \frac{{4{{\bar g}^2}\Omega \tilde \delta ({\kappa ^2} - {\omega ^2} + {{\tilde \delta }^2})}}{{{{({\kappa ^2} - {\omega ^2} + {{\tilde \delta }^2})}^2} + 4{\kappa ^2}{\omega ^2}}},
\end{equation}
and the effective mechanical damping rate,
\begin{equation}\label{gamaeff}
{\gamma _{eff}} = 2\gamma  + \frac{{8{{\bar g}^2}\Omega \tilde \delta \kappa }}{{{{({\kappa ^2} - {\omega ^2} + {{\tilde \delta }^2})}^2} + 4{\kappa ^2}{\omega ^2}}}.
\end{equation}
equations (\ref{ki}-\ref{gamaeff}) represents how the radiation pressure force modifies the dynamics of the mechanical mode.

Furthermore, $F_m(\omega)$ is the Fourier transform of the total force exerted on the mechanical mode, given by
\begin{eqnarray}\label{Ft}
{F_m}(\omega ) &= {\xi _p}(\omega ) + \frac{{\gamma  - i\omega }}{\Omega } {\xi _x}(\omega ) + \frac{{2\bar g\tilde \delta }}{{{{(\kappa  - i\omega )}^2} + {{\tilde \delta }^2}}} \xi_P \\ \nonumber
 &+ \frac{{2\bar g\tilde \delta (\kappa  - i\omega )}}{{{{(\kappa  - i\omega )}^2} + {{\tilde \delta }^2}}} \xi_X,
\end{eqnarray}
Inserting the above equations into equation (\ref{dispspedef}) and using the following noise correlation functions in the frequency domain
\numparts
\begin{eqnarray}\label{noiseomega}
\resizebox{0.64\textwidth}{!}{$\langle {{\xi_X}(\omega ){\xi_X}(\omega ')} \rangle  =[\kappa_0 (2{\bar n}_{th}+1) + \kappa_{ex} (u-v)^2] \delta (\omega  + \omega '), $}\\
\resizebox{0.64\textwidth}{!}{$\langle {{\xi_P}(\omega ){\xi_P}(\omega ')} \rangle  = [\kappa_0 (2{\bar n}_{th}+1) + \kappa_{ex} (u+v)^2] \delta (\omega  + \omega '),$}\\
\langle {{\xi_X}(\omega ){\xi_P}(\omega ')} \rangle  = \langle {{\xi_P}(\omega ){\xi_X}(\omega ')} \rangle ^{*}  = i \kappa \delta (\omega  + \omega '),\\
\resizebox{0.64\textwidth}{!}{$\langle {{\xi _x}(\omega ){\xi _x}(\omega ')} \rangle  = \langle {{\xi _p}(\omega ){\xi _p}(\omega ')} \rangle  = \gamma (2{{\bar n}_c} + 1)\delta (\omega  + \omega '),$}\\
\langle {{\xi _x}(\omega ){\xi _p}(\omega ')} \rangle  = \langle {{\xi _p}(\omega ){\xi _x}(\omega ')} \rangle ^* = i\gamma \delta (\omega  + \omega '), \label{noiseomege}
\end{eqnarray}
\endnumparts
we obtain
\begin{eqnarray}\label{Sx}
{S_x}&(\omega ) = \frac{1}{{4\pi }}|\chi (\omega ){|^2}\{ \gamma (2{{\bar n}_c} + 1)\frac{{{\Omega ^2} + {\gamma ^2} + {\omega ^2}}}{{{\Omega ^2}}} \nonumber \\
 &+ \frac{{4{{\bar g}^2}{\kappa _0}(2{{\bar n}_{th}} + 1)({{\tilde \delta }^2} + {\kappa ^2} + {\omega ^2})}}{{{{({\kappa ^2} - {\omega ^2} + {{\tilde \delta }^2})}^2} + 4{\kappa ^2}{\omega ^2}}} \nonumber\\
 &+ \frac{{4{{\bar g}^2}[{\kappa _{ex}}{{\tilde \delta }^2}{{(u + v)}^2} + {\kappa _{ex}}({\kappa ^2} + {\omega ^2}){{(u - v)}^2}]}}{{{{({\kappa ^2} - {\omega ^2} + {{\tilde \delta }^2})}^2} + 4{\kappa ^2}{\omega ^2}}}\} .
\end{eqnarray}
It is well-known  that in the strong coupling regime ($2 \bar g > \kappa$) the displacement spectrum shows NMS \cite{Aspelmeyer}. The NMS occurs due to the fact that when the OM coupling is strong, instead of having two separate subsystems, i. e., an optical mode with frequency $\tilde{\delta}$ and a mechanical mode with frequency $\Omega$, the system consists of two mixed modes. The eigenfrequencies of the normal modes are given by the eigenvalues of the drift matrix $\bf A$. When $\gamma , \kappa \ll \Omega , \tilde{\delta} $  the eigenfrequencies are approximately 
\begin{equation}\label{normalmodes}
\omega _ \pm ^2 \simeq \frac{1}{2}[({{\tilde \delta }^2} + {\Omega ^2}) \pm \sqrt {({{\tilde \delta }^2} - {\Omega ^2})^2 + 16\bar g^2 \Omega \tilde \delta } ].
\end{equation}
The NMS can be observed as two well-resolved peaks in the displacement spectrum. The properties of the photonic BEC affects theses peaks, so the measurement of the displacement spectrum of the membrane provides information about the photonic BEC. The normal modes (\ref{normalmodes}) depends on the BEC parameters in two ways. First, the effective detuning $\tilde{\delta}$ is a function of $N_0 V_0$ due to the Bogoliubov dispersion and second, the effective OM coupling $\bar g$ depends on $N_0$ and $V_0$.

\begin{figure}[htbp]
\centering
\includegraphics[width=3in]{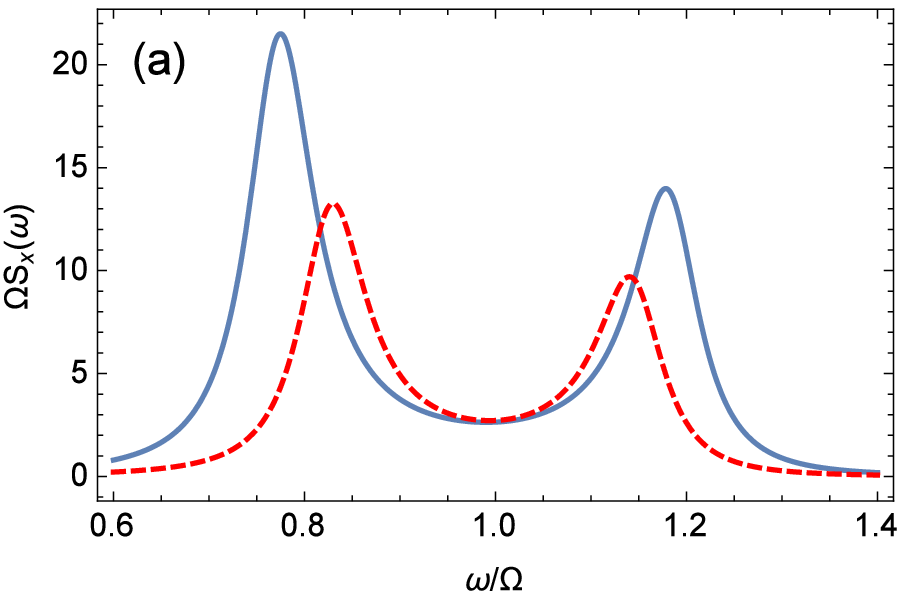}
\includegraphics[width=3in]{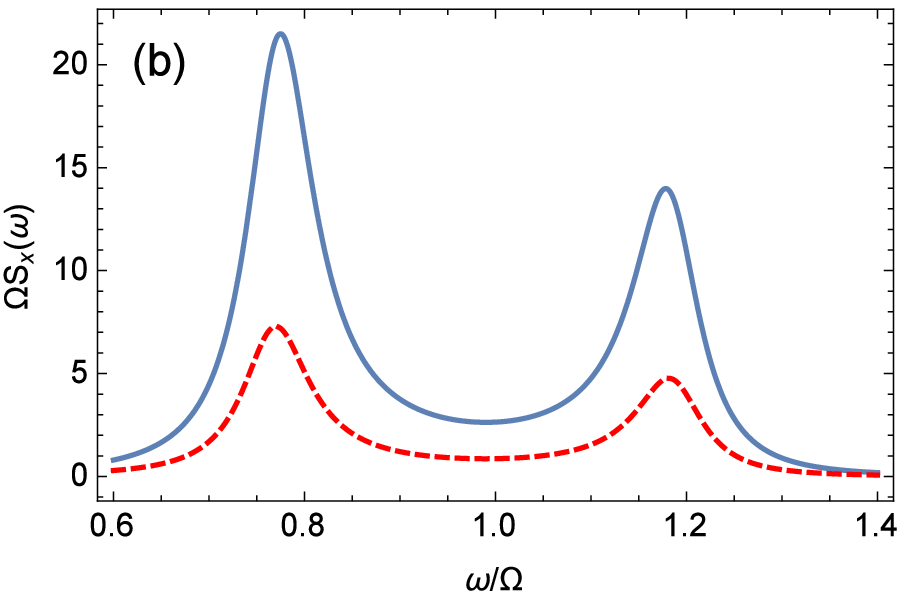}
\caption{(Color online) The normalized displacement spectrum of the mechanical mode versus the normalized frequency $\omega / \Omega$ for \textbf{(a) }$\zeta=4 \times 10^{-4}$ (blue solid line), $\zeta=10 \times 10^{-4}$ (red dashed line) and $k_B T=150 \hbar \omega_t $; \textbf{(b)} $k_B T=150 \hbar \omega_t$ (blue solid line), $k_B T=50 \hbar \omega_t$ (red dashed line) and $\zeta=4 \times 10^{-4}$ . The values of other parameters are $\tilde{\delta} = \Omega $, $\Omega = 7 \times 10^{-4} \omega_t, \kappa_{ex} = 10^{-5} \omega_t, \kappa_0 = 5 \kappa_{ex},  \gamma = 0.001 \kappa_{ex} , N_t = 10^6, g =4.2 \times 10^{-7} \omega_t$, and $  k_B T_m = 0.05 \hbar \omega_t. $ }
\label{fig2}
\end{figure}

In figure \ref{fig2}, we have plotted the normalized displacement spectrum of the membrane at resonance $\tilde{\delta}=\Omega$ versus the normalized frequency $\omega/ \Omega$ for different values of the parameter $\zeta$ (figure \ref{fig2}(a)) and the temperature $T$ (figure \ref{fig2}(b)). With increasing the photon-photon interaction strength while the temperature is fixed, the effective OM coupling is weakened and consequently, the splitting of the normal modes decreases (figure \ref{fig2}(a)). On the other hand, figure \ref{fig2}(b) shows that by decreasing the temperature while keeping the photon-photon interaction strength fixed, the heights of the two peaks of the spectrum decrease, although their positions remain almost unchanged. Actually, by decreasing the temperature the depletion of the photon BEC will decrease. However, since $N_0 \gg 1$ this change does not shift the peaks considerably but due to the smaller population of the Bogoliubov mode (see figure \ref{fig5}), it diminishes the heights of the peaks. It should be noted that the resonance condition itself depends on both $V_0$  and $T$. The numerical value of the parameters are chosen to be compatible with the experiment \cite{Klaers1}. For example, $k_B T/\hbar \omega_t = 150$ corresponds to the room temperature photonic BEC while the membrane is pre-cooled to about $100mK$. 

As can be seen from  equation (\ref{gamaeff}), it is evident that  the effective damping $\gamma_{eff}$ rate is larger than $2\gamma$ in the red-detuned regime ($\tilde{\delta} >0 $) so this extra damping can lead to the cooling of the membrane. The cooling occurs because, due to the finite lifetime of the photons, the radiation pressure force is non-conservative so that in the red-detuned regime it acts as a friction force for the mechanical element and causes the back-action cooling \cite{Aspelmeyer}. In addition, the back action cooling is more efficient when the light and the mechanical mode are in resonance. To quantify the cooling process the effective temperature $T_{eff}$ associated with the mechanical mode is defined by \cite{Clerk2}
\begin{equation}
\langle {{C^\dagger }C} \rangle {}_{ss} = \frac{1}{{{e^{\hbar {\Omega _{eff}}/{k_B}{T_{eff}}}} - 1}},
\end{equation}
where $\Omega_{eff} = \Omega_{eff}(\omega=\Omega)$. In addition in terms of the elements of the covariance matrix $\textbf{V}$, the steady-state phonon population is given by $\langle C^\dagger C \rangle_{ss}=(V_{33} + V_{44} -1)/2$ which can be calculated by solving the Lyapunov equation (\ref{lyap}). 

\begin{figure}[htbp]
\centering
\includegraphics[width=3in]{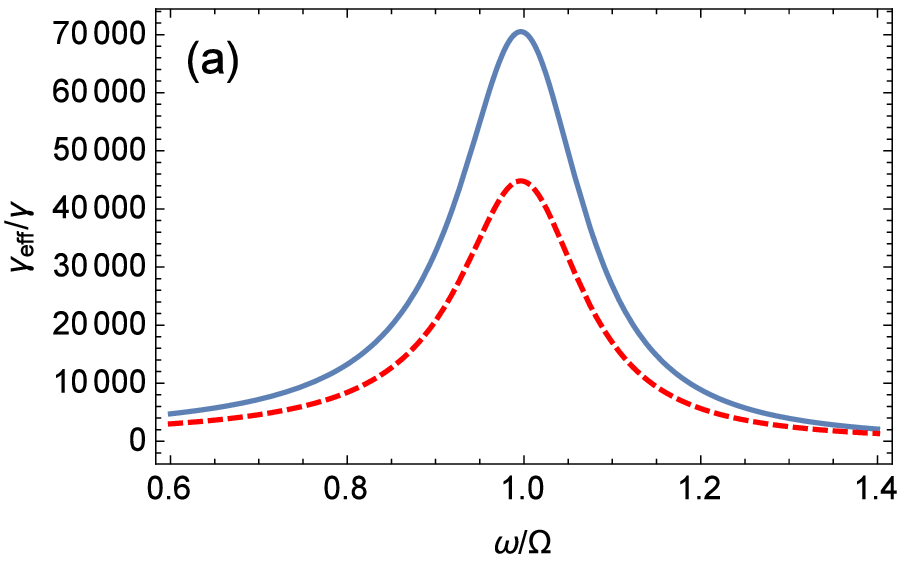}
\includegraphics[width=3in]{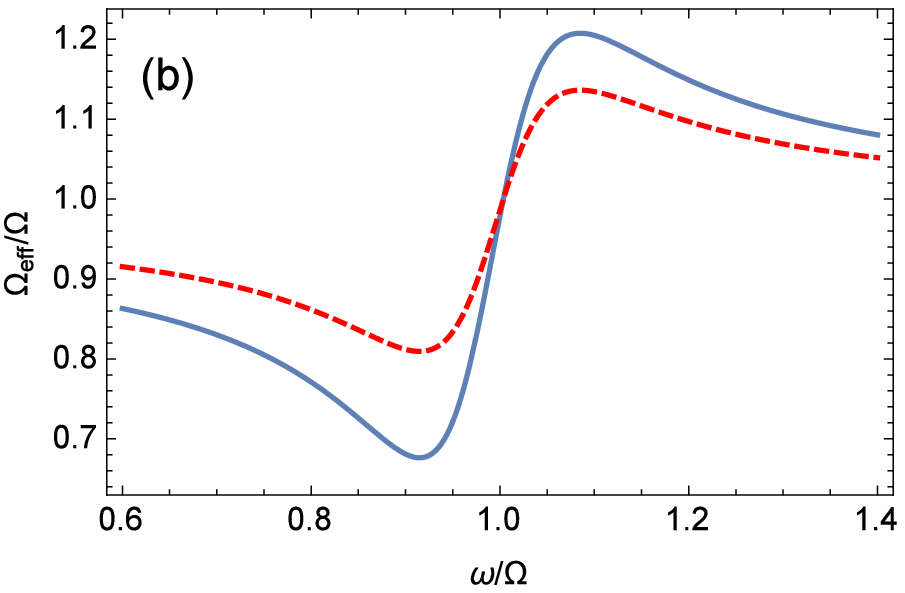}
\caption{(Color online)  \textbf{(a)} The normalized effective mechanical damping rate and \textbf{(b)} the normalized effective mechanical frequency versus the normalized frequency $\omega / \Omega$ for $\zeta=4 \times 10^{-4}$ (blue solid line) and $\zeta=10 \times 10^{-4}$ (red dashed line). Here, we have set $k_B T=150 \hbar \omega_t $. The values of other parameters are the same as those in figure \ref{fig2}.   }
\label{fig3}
\end{figure}

 Figures \ref{fig3}(a) and \ref{fig3}(b) illustrate, respectively, the normalized effective mechanical damping and normalized effective mechanical frequency versus $\omega / \Omega$ for two values of the interaction parameter $\zeta$ assuming the resonance condition and keeping temperature fixed. Figure \ref{fig3}(a) shows that $\gamma _{eff}$ is considerably greater than $\gamma$ and it decreases by increasing the photon-photon interaction strength. The effective mechanical frequency decreases with increasing $\zeta$ (figure \ref{fig3}(b)). Since the figures are plotted at resonance, these changes arise from the decrease in the effective OM coupling.

\begin{figure}[htbp]
\centering
\includegraphics[width=3in]{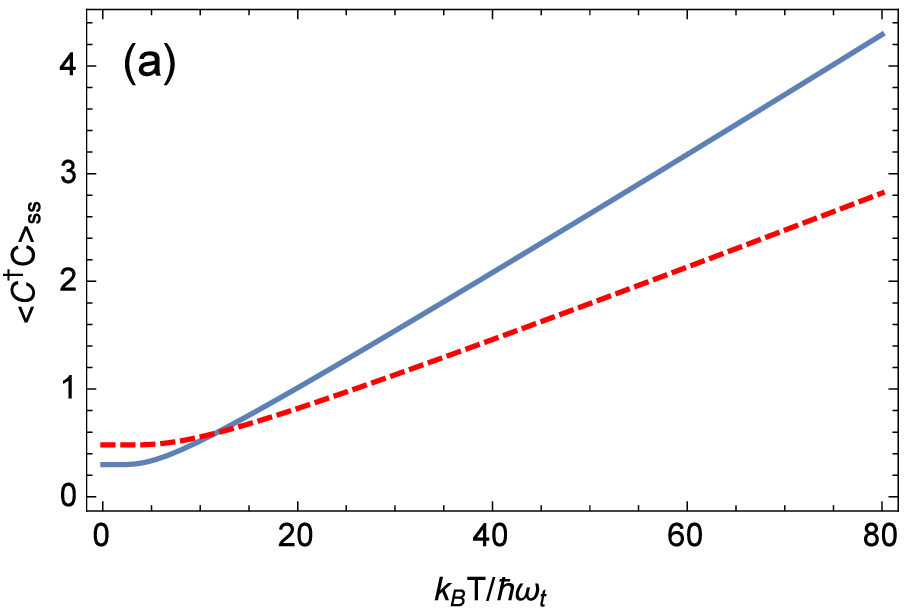}
\includegraphics[width=3in]{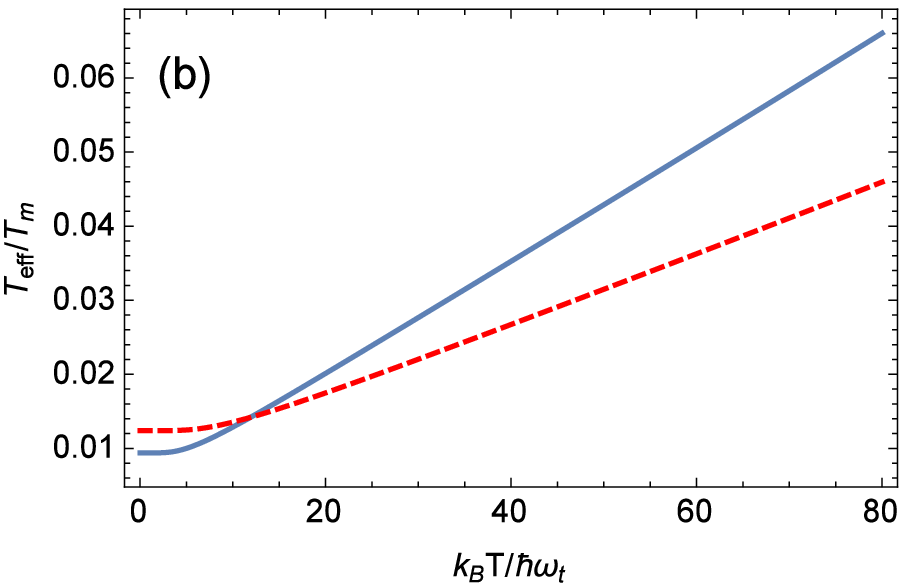}
\caption{(Color online) (a) The steady-state phonon occupation and (b) the normalized effective temperature versus the normalized temperature of the BEC of photon for (a) $\zeta=4 \times 10^{-4}$ (blue solid line) and $\zeta=10 \times 10^{-4}$ (red dashed line). The values of other parameters are the same as those in figure \ref{fig2}.   }
\label{fig4}
\end{figure}

\begin{figure}[htbp]
\centering
\includegraphics[width=3in]{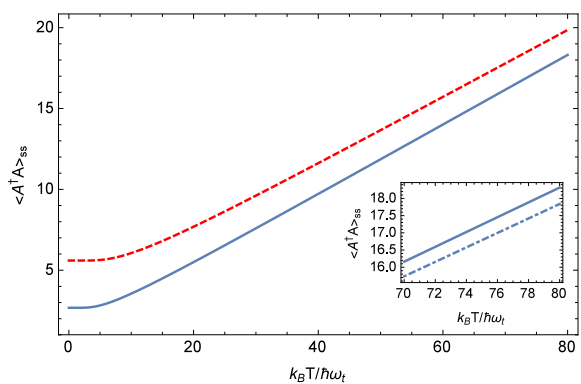}
\caption{(Color online) The steady-state population of the excited mode of the cavity versus the temperature for $\zeta=4 \times 10^{-4}$ (blue solid line), $\zeta=10 \times 10^{-4}$ (red dashed line) and $ g = 8.4 \times 10^{-7} \omega_t $ . The values of other parameters are the same as those in figure \ref{fig2}. In the inset, the dot-dashed blue line represents the corresponding photon population in the absence of OM coupling, $g=0$. }
\label{fig5}
\end{figure}

The steady-state phonon number and the associated effective temperature are, respectively, plotted in figures \ref{fig4}(a) and  \ref{fig4}(b) versus the photonic BEC temperature for two values of the $\zeta$. It is obvious that at lower temperatures the thermal noise is small and the cooling process is more efficient such that for very low temperatures the ground state cooling, $\langle C^\dagger C\rangle_{ss}<1$,  is even possible (figure \ref{fig4}(a)). On the other hand,  by increasing the two-body interaction potential, in spite of the reduction of the effective mechanical damping rate (figure \ref{fig3}(a)), the phonon occupation number decreases. This result can be understood by noting that when the system operates in the strong OM coupling regime, the amplification of the OM coupling leads to an increase of the phonon occupancy \cite{Aspelmeyer}. However, in the system under consideration, increasing the photon-photon repulsion reduces the effective OM coupling, and thus the phonon occupancy decreases. Thus, the mechanical cooling is more efficient in the interacting photon BEC, but as seen in figure \ref{fig4}, this is not the case when the photon BEC temperature is very low. The reason is that at low temperatures the thermal photon number $\bar n _{th}$  is negligible and in this limit the effect of input noise correlations (equation \ref{bnoiset}) (see also diffusion matrix in section \ref{secmotion}) becomes significant. The photon-photon interaction amplifies the effects of this noise which limits the cooling efficiency.  The lower limit of the effective temperature ($\simeq 0.01 T_m$) corresponds to the temperature about $1 mK$ for the experimental setup introduced in \cite{Klaers1}.

To end up this section, we consider the depletion of the BEC of photons. For this purpose, we have illustrated the temperature dependence of the steady-state number of photons in the excited mode of the cavity (figure \ref{fig5}), which in terms of the covariance matrix elements is given by $\langle A^\dagger A \rangle_{ss}=(V'_{11} + V'_{22} -1)/2$. As is seen, increasing the photon-photon interaction as well as the temperature leads to more population of the excited mode. The OM coupling also slightly increases the photon condensate depletion (see inset of figure \ref{fig5}). However, as is evident, the amount of depletion induced by both the photon-photon and OM interactions is negligibly small compared to $N_t$. Thus, one can safely neglect the effect of photon depletion in $N_0$ .

 \section{Output intensity and quadrature squeezing spectra} \label{secinten}

The other measurable quantities of the system which can be examined to obtain information about the BEC of photons in the cavity and its OM interaction are the output intensity and quadrature squeezing spectra. In this section, we calculate these spectra to explore how they depend on the features of the photonic BEC. The output intensity spectrum is defined as the Fourier transform of the output field correlation function \cite{Clerk2}
 \begin{equation}\label{soutdef}
{S_{out}}(\omega ) = \frac{1}{{2\pi }}\int\limits_{ - \infty }^\infty  {d\tau \langle {A_{out}^\dagger (t + \tau ){A_{out}}(t)} \rangle {e^{ - i\omega \tau }}} .
\end{equation}
where the output field $A_{out}$ is related to the input field $A_{in}$ via $A_{out} = \sqrt{2\kappa_{ex}} A - A_{in}$. Considering the correlation functions (\ref{noiseomega}-$e$), the symmetric output intensity spectrum is given by $ {S^s_{out}}(\omega ) = [{S_{out}}(\omega ) + {S_{out}}(-\omega )]/2$ with ${S_{out}}(\omega ) =\langle {A_{out}^\dagger}(\omega) A_{out}(-\omega) \rangle$.
Solving the equation of motion (\ref{matrixeq}) in the frequency domain, we get 
 \begin{equation}\label{Bo}
 B(\omega ) = {\alpha _1}(\omega )\xi_B + {\alpha _2}(\omega )\xi_B^\dagger + {\alpha _3}(\omega ) {\xi _c} + {\alpha _4}(\omega ) \xi _{c}^\dagger ,
 \end{equation}
 where
 \numparts
\begin{eqnarray}\label{alphas}
 {\alpha _1}(\omega ) &=& {d^{ - 1}}\{ [{(\gamma  - i\omega )^2} + {\Omega ^2}][i(\tilde \delta  + \omega ) - \kappa ] \nonumber\\
&-& 2i{{\bar g}^2}\tilde \delta \Omega \} ,\\
{\alpha _2}(\omega ) &=& {d^{ - 1}}( - 2i{{\bar g}^2}\Omega ),\\
 {\alpha _3}(\omega ) &=& {d^{ - 1}}[i\bar g(\tilde \delta  + i\kappa  + \omega )(\Omega  + i\gamma  + \omega )],\\
 {\alpha _4}(\omega ) &=& {d^{ - 1}}[ - i\bar g(\tilde \delta  + i\kappa  + \omega )(\Omega  - i\gamma  + \omega )];
\end{eqnarray}
\endnumparts
with
\begin{equation}
d =  - [{(\gamma  - i\omega )^2} + {\Omega ^2}][{(\kappa  - i\omega )^2} - {{\tilde \delta }^2}] + 4{{\bar g}^2}\tilde \delta \Omega .
\end{equation}
By applying the Bogoliubov transformation (\ref{bogo}), the relation $B^{\dagger}(\omega) = [B(-\omega)]^{\dagger}$, and the noise correlation functions (\ref{noiseomega}-$e$) we obtain, after some algebra, the output intensity spectrum as follows
 \begin{equation}\label{s123}
{S_{out}}(\omega ) = 2\kappa_{ex} \{ {u^2}{\beta _1}(\omega ) + {v^2}{\beta _2}(\omega ) + 2uv{\mathop{\rm Re}\nolimits} [{\beta _3}(\omega )]\} ,
\end{equation}
where the functions $\beta _i$ are defined as 
\numparts
\begin{eqnarray}\label{betas}
{\beta _1}(\omega ) &= [2\kappa_{ex} v^2+2\kappa_0 {{\bar n}_{th}}] |{\alpha _1}(-\omega ){|^2} \nonumber \\
&+ [2\kappa_{ex} u^2+2\kappa ({{\bar n}_{th}} + 1)] |{\alpha _2}(-\omega ){|^2} \nonumber\\ 
&- 4\kappa_{ex} uv Re[\alpha(-\omega)^* \alpha_2(-\omega)] \nonumber \\
&+ 2\gamma {{\bar n}_c}|{\alpha _3}(-\omega ){|^2} + 2\gamma ({{\bar n}_c} + 1)|{\alpha _4}(-\omega ){|^2},\\
{\beta _2}(\omega ) &= [2\kappa_{ex} u^2+2\kappa ({{\bar n}_{th}} + 1)] |{\alpha _1}(  \omega ){|^2} \nonumber \\
&+ [2\kappa_{ex} v^2+2\kappa_0 {{\bar n}_{th}}] |{\alpha _2}( \omega ){|^2} \nonumber\\ 
 &- 4\kappa_{ex} uv Re[\alpha(\omega) \alpha_2(\omega)^*] \nonumber \\
 &+ 2\gamma ({{\bar n}_c} + 1)|{\alpha _3}( \omega ){|^2} + 2\gamma {{\bar n}_c}|{\alpha _4}(  \omega ){|^2},\\
{\beta _3}(\omega ) &= [2{\kappa _{ex}}{u^2} + 2{\kappa _0}({{\bar n}_{th}} + 1)]{\alpha _1}(\omega ){\alpha _2}( - \omega ) \nonumber \\ 
 &+ [2{\kappa _{ex}}{v^2} + 2{\kappa _0}{{\bar n}_{th}}]{\alpha _2}(\omega ){\alpha _1}( - \omega ) \nonumber \\ 
& - 2{\kappa _{ex}} uv [{\alpha _1}(\omega ){\alpha _1}( - \omega ) + {\alpha _2}(\omega ){\alpha _2}( - \omega )] \nonumber \\ 
 &+ 2\gamma ({{\bar n}_c} + 1){\alpha _3}(\omega ){\alpha _4}( - \omega ) \nonumber \\
 & + 2\gamma {{\bar n}_c}{\alpha _3}( - \omega ){\alpha _4}(\omega ).
\end{eqnarray}
\endnumparts

Furthermore, to investigate the squeezing properties of the output field we calculate the quadrature noise spectrum of the output field defined by \cite{Giovannetti},
\begin{equation}\label{Sphi}
\resizebox{0.65\textwidth}{!}{${S_\phi }(\omega ) = \frac{1}{{4\pi }}\int_{ - \infty }^\infty  {d\omega {e^{i(\omega  + \omega ')t}}\langle {{X_\phi }(\omega ){X_\phi }(\omega ') + {X_\phi }(\omega '){X_\phi }(\omega )} \rangle } , $}
 \end{equation}
 with ${X_\phi }(\omega ) = {e^{ - i\phi }}{A_{out}}(\omega ) + {e^{i\phi }}A_{out}^\dagger (\omega )$. When ${S_\phi }(\omega ) <1$ the output field is squeezed. By minimizing ${S_\phi }(\omega ) $ with respect to $\phi$, the optimized quadrature squeezing spectrum is obtained as 
 \begin{equation}\label{Sopt}
 S_{opt}(\omega)=S_{out}^s(\omega)+\mathcal{C}^s_{AA^{\dagger}}(\omega)-2 | \mathcal{C}^s_{AA}(\omega) |^2,
 \end{equation}
 where the optimum $\phi$ satisfies $ e^{2i\phi _{opt}} = - \frac{\mathcal{C}^s_{AA^{\dagger}}(\omega)}{| \mathcal{C}^s_{AA}(\omega) | } $. In addition, the symmetric functions $\mathcal{C}^s_{AA^{\dagger}}(\omega)$ and $\mathcal{C}^s_{AA}(\omega)$ are given by
\numparts
\begin{eqnarray}\label{caas}
 {\cal C}^{s}_{AA^{\dagger}}(\omega)&=\frac{1}{2} {[\cal C}_1(\omega)-{\cal C}_2(\omega)+{\cal C}_1(-\omega)-{\cal C}_2(-\omega)] \nonumber \\
 &+1, \\
 {\cal C}^s_{AA}(\omega)&=\frac{1}{2} {[\cal C}_3(\omega)-{\cal C}_4(\omega)+{\cal C}_3(-\omega)-{\cal C}_4(-\omega)],
\end{eqnarray}
\endnumparts
with
\numparts
\begin{eqnarray}\label{cis}
 {{\cal C}_1}(\omega ) &= 2\kappa_{ex} \{ {u^2}{\beta _2}(\omega ) + {v^2}{\beta _1}(\omega ) + 2uv{\mathop{\rm Re}\nolimits} [{\beta _3}(\omega )]\} ,\\
{{\cal C}_2}(\omega ) &= 4 \kappa_{ex} Re[\alpha_1(\omega)],\\
{{\cal C}_3}(\omega ) &=2\kappa_{ex} \{ {u^2}{\beta _3}(\omega ) + {v^2}{\beta _3}{(\omega )^*} + uv[{\beta _1}(\omega ) + {\beta _2}(\omega )] \} ,\\
 {{\cal C}_4}(\omega ) &= 2{\kappa _{ex}}[{u^2}{\alpha _2}( - \omega ) - {v^2}{\alpha _2}{(\omega )^*} - uv{\alpha _1}( - \omega ) \nonumber \\
&+ uv{\alpha _1}{(  \omega )^*}].
\end{eqnarray}
\endnumparts

\begin{figure}[htbp]
\centering
\includegraphics[width=3in]{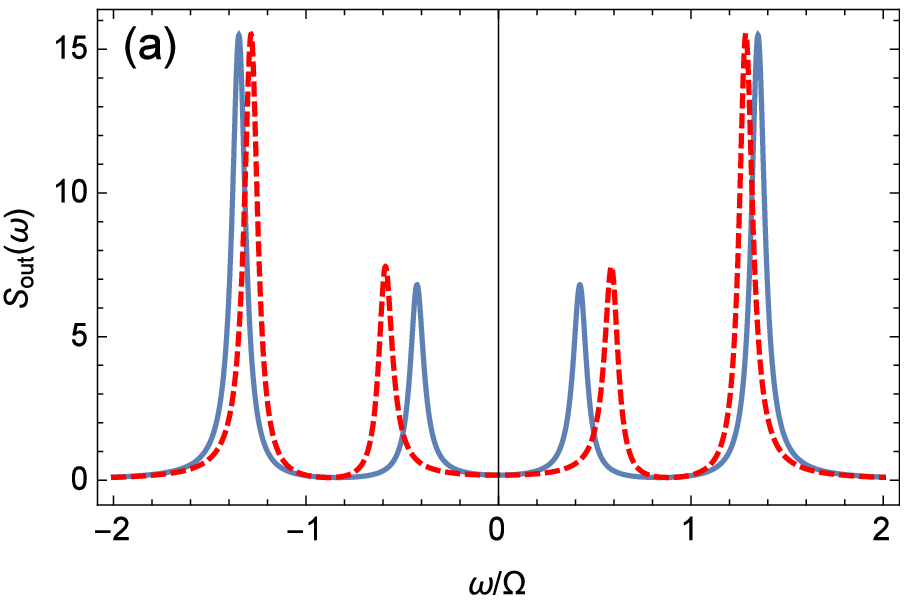}
\includegraphics[width=3in]{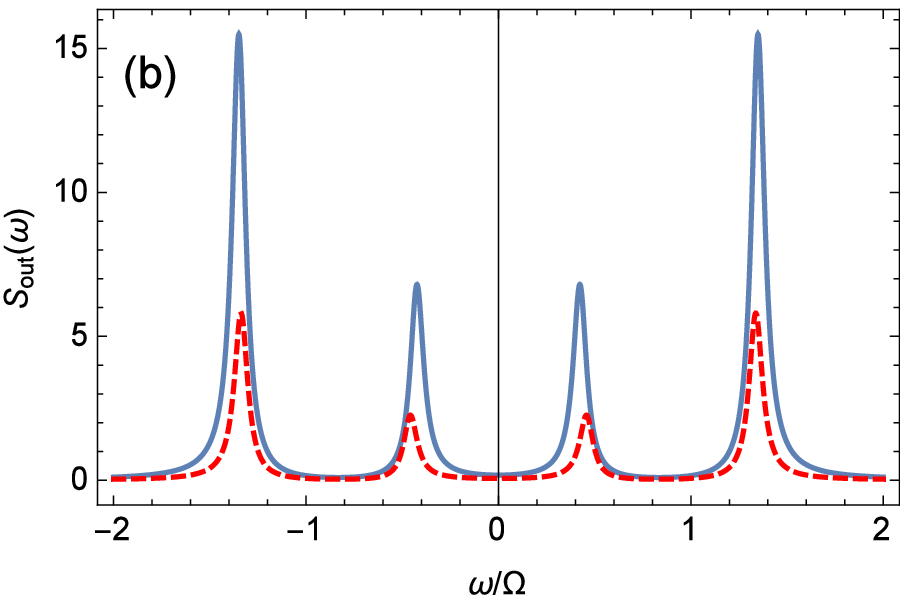}
\caption{(Color online)  The output-field intensity spectrum versus the normalized frequency $\omega /\Omega$ for $ g = 8.4 \times 10^{-7} \omega_t $ and \textbf{(a)} $\zeta=4 \times 10^{-4}$ (blue solid line), $\zeta=10 \times 10^{-4}$ (red dashed line) and $k_B T=150 \hbar \omega_t $; \textbf{(b)} $k_B T=150 \hbar \omega_t$ (blue solid line), $k_B T=50 \hbar \omega_t$ (red dashed line) and $\zeta=4 \times 10^{-4}$. The values of other parameters are the same as those in figure \ref{fig2}.  }
\label{fig6}
\end{figure}

\begin{figure}[htbp]
\centering
\includegraphics[width=3in]{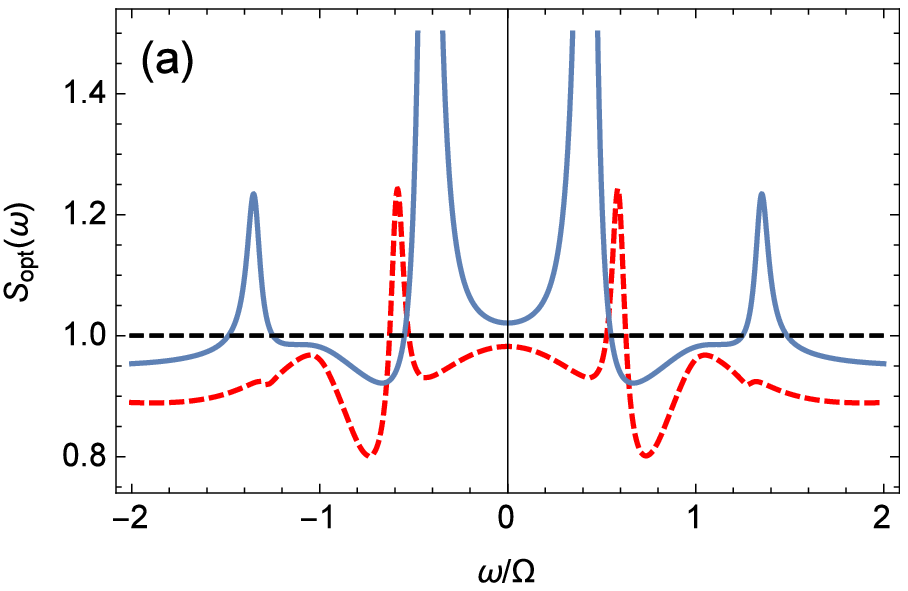}
\includegraphics[width=3in]{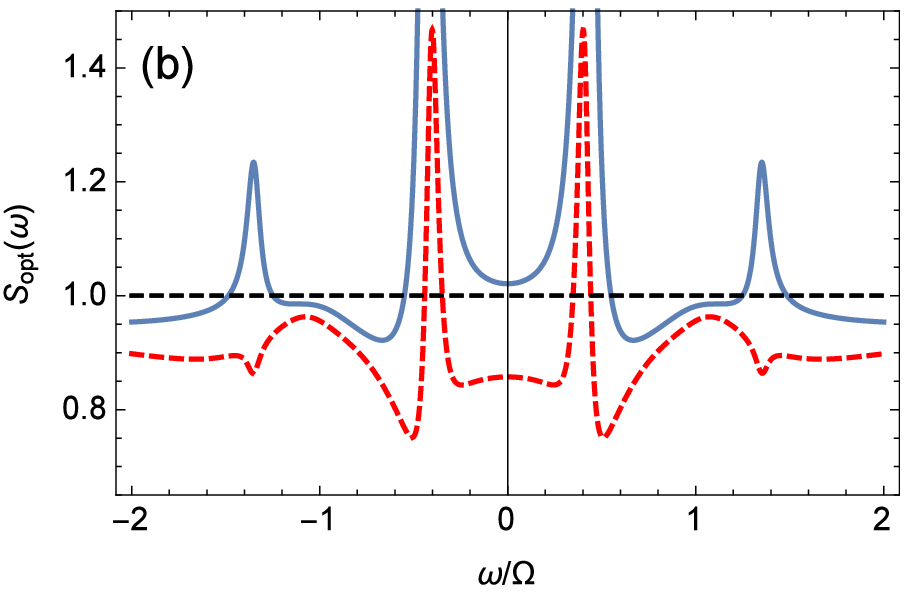}
\includegraphics[width=3in]{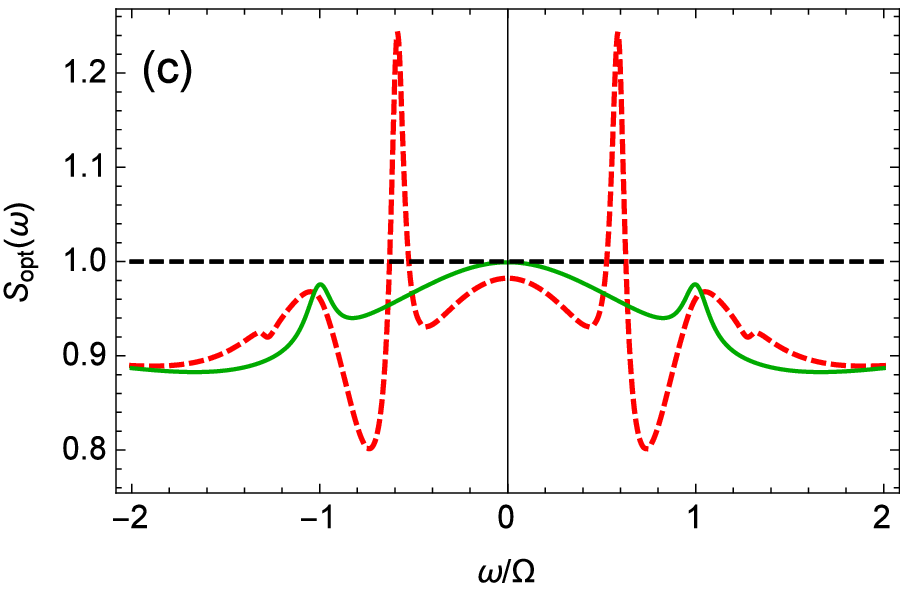}
\caption{(Color online) The output-field quadrature squeezing spectrum versus the normalized frequency $\omega / \Omega$ for \textbf{(a)} $\zeta=4 \times 10^{-4}$ (blue solid line), $\zeta=10 \times 10^{-4}$ (red dashed line), $k_B T=150 \hbar \omega_t $ and $ g = 8.4 \times 10^{-7} \omega_t $; \textbf{(b)} $k_B T=150 \hbar \omega_t$ (blue solid line), $k_B T=50 \hbar \omega_t$ (red dashed line), $\zeta=4 \times 10^{-4}$ and $ g = 8.4 \times 10^{-7} \omega_t $; and \textbf{(c)} $ g = 8.4 \times 10^{-7} \omega_t $ (red dashed line), $ g = 0$ (green solid line), $\zeta=10 \times 10^{-4}$ and $k_B T=150 \hbar \omega_t$. The values of other parameters are the same as those in figure \ref{fig2}.   }
\label{fig7}
\end{figure}

The output intensity and the quadrature squeezing spectra are plotted in figures \ref{fig6} and \ref{fig7}, respectively. Increasing the photon-photon interaction strength reduces the NMS in the output intensity  spectrum via diminishing the OM coupling and of course, increases the squeezing of the output field due to amplification of the parametric interaction (figure \ref{fig7}(a)). In addition, the change of the temperature of the photonic BEC does not affect the splitting of the normal modes considerably. However, with decreasing the temperature, the thermal noise is reduced and thus, the depletion of the photonic BEC will be decreased (figure \ref{fig5}) which cause to the attenuation of the output intensity (figure \ref{fig6}(b)) and enhancement of the squeezing of the output field (figure \ref{fig7}(b)). Figure \ref{fig7}(c) shows that in the absence of the OM coupling of the photonic BEC to the membrane ($g=0$), the microcavity output field exhibits quadrature squeezing even at room temperature. With increasing the OM coupling the output-field quadrature  squeezing is strengthened.

\section{The photon-phonon Entanglement}\label{secentan}

One of the most important quantum features of the OM systems is that the radiation pressure can lead to the steady-state entanglement between the subsystems. The entanglement measure which is usually used to quantify the entanglement of the bimodal Gaussian state is the logarithmic negativity \cite{Genes2}. Here, we are interested in the entanglement between the excited modes of the photonic BEC and the mechanical modes in the system under consideration and the logarithmic negativity is convenient for our purpose.

The logarithmic negativity is defined as \cite{Vidal}
\begin{equation}
 {\cal {E_N}} =\max \{ 0,-\ln 2{\eta ^ - }\} , 
\end{equation}
 where
\begin{equation}\label{eta}
{\eta ^ - } = \frac{1}{{\sqrt 2 }}{[\Sigma{({\bf{V'}})}  - \sqrt {\Sigma {({\bf{V'}})}  - 4\det {\bf{V'}}} ]^{1/2}},
\end{equation}
is the lowest symplectic eigenvalue of the partial transpose of the associated covariance matrix (\ref{Vp}). Here, $\Sigma (\bf{V}')= \det{\bf {V}_1} + \det{\bf {V}_2} - 2\det{\bf {V}_{\cal {C}}} $, where $\bf {V}_1$, $\bf {V}_2$ and $\bf {V}_{\cal {C}}$ are $2\times 2$  block matrices defined by 
\begin{equation}
{\bf{V'}} = \left( {\begin{array}{*{20}{c}}
{{{\bf{V}}_1}}&{{{\bf{V}}_{\cal {C}}}}\\
{{\bf{V}}_{\cal {C}}^\dagger }&{{{\bf{V}}_2}}
\end{array}} \right).
\end{equation}
Therefore, by solving the Lyapunov equation (\ref{lyap}) the covariance matrix $\bf {V}'$ and so the logarithmic negativity can be calculated.

\begin{figure}[htbp]
\centering
\includegraphics[width=3in]{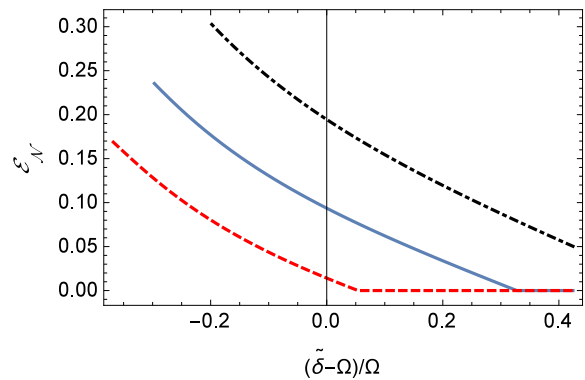}
\caption{(Color online) The logarithmic negativity versus the normalized detuning $(\tilde{\delta} - \Omega)/\Omega$ for $\zeta=5 \times 10^{-4}$ (red dashed line), $\zeta=4 \times 10^{-4}$ (blue solid line), $\zeta=3 \times 10^{-4}$ (black dot-dashed line), $k_B T=1.5 \hbar \omega_t$, and $ g = 8.4 \times 10^{-7} \omega_t $. The values of other parameters are the same as those in figure \ref{fig2}.  }
\label{fig8}
\end{figure}

\begin{figure}[htbp]
\centering
\includegraphics[width=3in]{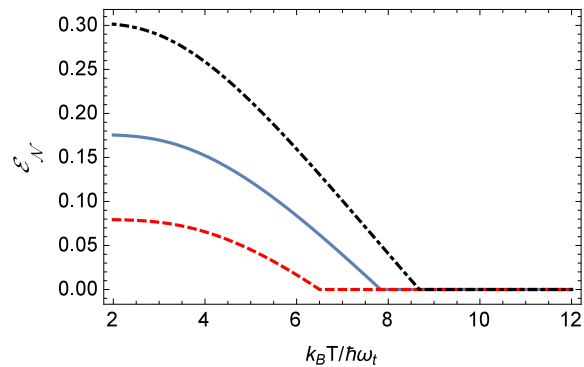}
\caption{(Color online) The logarithmic negativity versus the normalized temperature of the photon condensate for $\zeta=5 \times 10^{-4}$ (red dashed line), $\zeta=4 \times 10^{-4}$ (blue solid line), $\zeta=3 \times 10^{-4}$ (black dot-dashed line), $\tilde{\delta} - \Omega = -0.2\Omega $, and $ g = 8.4 \times 10^{-7} \omega_t $. The values of other parameters are the same as those in figure \ref{fig2}.   }
\label{fig9}
\end{figure}

In figure (\ref{fig8}) the logarithmic negativity is plotted versus the normalized detuning $(\tilde{\delta} - \Omega)/\Omega$ for different values of the parameter. The figure shows that the degree of photon-phonon entanglement decreases when the photon-photon scattering becomes stronger. This is due to the reduction of the effective OM coupling $\bar{g}$,. In addition, for the system to be in an entangled state, the effective detuning should be adjusted close to the instability region. Furthermore, the plot of the logarithmic negativity versus the BEC temperature (figure (\ref{fig9})) shows that there is a threshold temperature that above it the thermal noise prevents the system to be in an entangled state. As can be seen from figure (\ref{fig9}) the threshold temperature depends on the nonlinearity induced by the photon-photon interaction; the weaker the nonlinearity of the photon gas, the higher is the threshold temperature. 

\section{Conclusions} \label{seccon}
To summarize, we have considered a system consisting of a photon BEC much below the threshold and an oscillating membrane in a micro-cavity. We have investigated the OM dynamics of the system by applying the Bogoliubov approximation to the photonic BEC. In this approximation, the macroscopic occupation of the ground state of the cavity not only leads to a strong coupling between the collective excitation mode of the photonic BEC and the mechanical motion but also causes a strong parametric interaction in the photon gas. 

The study of the radiation pressure back action cooling shows that by decreasing the temperature of the photonic BEC, it is possible to cool down the mechanical element to its ground state. Although lower effective temperature can be achieved when the photon-photon interaction is stronger, but it is not the case for low temperatures since further cooling is prevented by the correlation of the input noise. Furthermore, the NMS can be observed in the displacement spectrum of the membrane as well as in the output intensity spectrum which decreases with increasing the nonlinearity induced by the photon-photon interaction.  

We have examined the field quadrature noise spectrum which demonstrates that the output field from the microcavity can be quadrature squeezed. Based our results, the amount of squeezing is enhanced when the photon-photon repulsion and the OM coupling increases and the BEC temperature decreases. We have also shown that a stationary entanglement can be established between the membrane and the collective excited mode of the photonic BEC when the temperature is below a threshold value and the nonlinear photon-photon interaction is weak enough.

\end{document}